\documentclass[12pt]{iopart}
\usepackage{graphicx}
\usepackage{dcolumn}
\usepackage{bm}
\begin{document}

\title{Ultra-High Energy Cosmic Rays from Quasar Remnants}
\author{Claudia Isola$^{1,2}$, G{\"u}nter Sigl$^{2}$, Gianfranco Bertone$^{2,3}$}

\address{1 \ Dipartimento di Fisica, Universit\`a di Cagliari and INFN, Sezione di Cagliari, ITALY.
Strada provinciale per Sestu Km 0.700 09042 Monserrato (Cagliari)}

\address{2 \ GReCO, Institut d'Astrophysique de Paris, C.N.R.S., 98 bis boulevard
Arago, F-75014 Paris, France}

\address{3 \ NASA/Fermilab Astrophysics Center, 
Fermi National Accelerator Laboratory, Batavia IL, 60510-0500}

\ead{isola@iap.fr, sigl@iap.fr, bertone@trovatore.fnal.gov }

\begin{abstract}
One of the models recently proposed to explain the origin of the
ultra high energy cosmic rays assumes that these particles may be
accelerated by the electromotive force around presently inactive
quasar remnants. We study predictions for large and small scale
ultra high energy cosmic ray arrival direction anisotropies in a
scenario where the particles are injected with a mono-energetic
spectrum by a discrete distribution of such sources. We find that
known quasar remnants are typically distributed too anisotropically
to explain the isotropic ultra high energy cosmic ray flux except
in the unrealistic case where extragalactic magnetic fields of
$\simeq 0.1\mu$G extend over many Mpc.
\end{abstract}



FERMILAB-Pub-03/402-A, MADPH-03-1337

\submitto{Journal of Cosmology and Astroparticle Physics, JCAP}

\maketitle

\section{Introduction}

Over the last few years the detection of several giant air showers,
either through ground based detectors~\cite{haverah,agasa}
or fluorescence telescopes ~\cite{fe,hires}, have
confirmed the arrival of ultra high energy cosmic-rays (UHECRs)
with energies up to a few hundred EeV (1 EeV $\equiv 10^{18}\,$eV).
Their existence poses a serious challenge and is
currently subject of much theoretical research
as well as experimental efforts (for recent reviews see
\cite{reviews,bs-rev,school}).

The problems encountered in trying to explain UHECRs in terms of
``bottom-up'' acceleration mechanisms have been well-documented
in a number of studies (e.g., Refs.~\cite{hillas-araa,ssb,norman}).
In summary, apart from energy draining interactions in the source,
the maximal UHECR energy is limited by the product
of the accelerator size and the strength of the magnetic field.
According to this criterion it turns out that it is
very hard to accelerate protons and heavy nuclei
up to the observed energies, even for the most powerful
astrophysical objects
such as radio galaxies and active galactic nuclei.

In addition, nucleons above
$\simeq50\,$EeV suffer heavy energy losses due to
photo-pion production on the cosmic microwave background (CMB)
--- the Greisen-Zatsepin-Kuzmin (GZK) effect~\cite{gzk} ---
which limits the distance to possible sources to less than
$\simeq100\,$Mpc~\cite{stecker}. Heavy nuclei at these energies
are photo-disintegrated in the CMB within a few Mpc~\cite{heavy}.
Unless the sources are strongly clustered in our
local cosmic environment, a drop, often called the ``GZK cut-off'' in
the spectrum above $\simeq50\,$EeV is therefore expected~\cite{bbo},
even if injection spectra extend to much higher energies.
However, the existence of the latter is not established yet
from the observations~\cite{mbo}. In fact, whereas a cut-off
seems consistent
with the few events above $10^{20}\,$eV recorded by the fluorescence
detector HiRes~\cite{hires}, it is not compatible with the
8 events (also above $10^{20}\,$eV) measured by the AGASA ground
array~\cite{agasa}. The solution of this problem
may have to await the completion of the Pierre Auger project~\cite{auger}
which will combine the two complementary detection techniques
adopted by the aforementioned experiments.

Adding to the problem, there are no obvious astronomical counterparts
to the detected UHECR events
within $\simeq 100\,$Mpc of the Earth~\cite{elb-som,ssb}. At the same
time, no significant large-scale anisotropy has been observed in
UHECR arrival directions above $\simeq10^{18}\,$eV,
whereas there are strong hints for small-scale clustering:
The AGASA experiment has observed four doublets and one triplet within
$2.5^\circ$ out of a total of 57 events detected above 40 EeV~\cite{agasa}.
When combined with three other ground array experiments, these numbers increase
to at least eight doublets and two triplets within $4^\circ$~\cite{uchihori}.
This clustering has a chance probability of less than
$1\%$ in the case of an isotropic distribution.

There are currently two possible explanations of these
experimental findings. {In the first one, which assumes negligible
magnetic deflection, most of the sources would have to be at
cosmological distances which would explain the absence of nearby
counterparts and the apparent isotropy would indicate that many
sources contribute to the observed flux, where a subset of
specially powerful sources would explain the small-scale
clustering~\cite{tt}. This scenario predicts the confirmation of
a GZK cutoff. The second scenario is more realistic and takes into
account the likely existence of large scale intervening magnetic
fields correlated with the large scale galaxy distribution. In
this case magnetic deflection could be considerable even at the
highest energies and the observed UHECR flux could be dominated by
relatively few sources within about 100 Mpc. Here, large scale
isotropy could be explained by considerable angular deflection
leading to diffusion up to almost the highest energies and magnetic
lensing~\cite{hmrs} could contribute to the small scale clustering.

A possible way of explaining the origin of these particles is to
assume a source population associated with quasar remnants
currently inactive in the visible spectrum~\cite{bg}. If these
sources are sufficiently numerous, this would explain large scale
isotropy, whereas a few nearby sources could explain the small
scale clustering. In this scenario, UHECRs would be accelerated at
the nuclei of those nearby dead quasars, whose underlying
super-massive black holes are sufficiently spun-up to provide the
necessary electromotive force. Hints of a possible correlation
between some of these dormant quasars and the arrival direction of
cosmic rays above $4\times10^{19}\,$eV may support this
scenario~\cite{tbhl}.

In the present study we perform numerical simulations in order to
test a sample of candidate dead quasars as possible sources for
the ultra high energy events. We first considered  the case of an
all-pervading turbulent magnetic field of 1nG strength, which
corresponds to the primordial field necessary to explain galactic
magnetic fields by adiabatic compression~\cite{bt_review}. We also
performed simulations with a relatively strong magnetic field of $0.1\mu$G,
either following a pancake like structure mimicking the Local
Supercluster, or homogenous up to the farthest sources. All these
cases are consistent with Faraday rotation upper
limits~\cite{blasi}, although the strong field case may not be
very realistic.

The paper is organised as follows: in Sec.II we outline the scenario
underlying our numerical simulations which are briefly described in
Sec.III. In Sec.IV we introduce the statistical quantities used for
comparison with the data. In Sec.V we present our results and
we conclude in Sec.VI.

\section{The Quasar Remnant Scenario}

Boldt and Ghosh~\cite{bg} suggested that supermassive black holes,
present in the center of normal galaxies~\cite{magorrian},
could be the principal sources of particles at the highest energies.
As stressed in the introduction, the lack of observed
counterparts to the highest energy events within an acceptable
distance suggests the idea of sources currently inactive in the
visible spectrum.
These objects should be able to accelerate UHECRs
and appear dormant at the same time. Quasars
remnants or dormant AGNs (active in earlier phases),
with underlying supermassive black holes,
are among the astronomical objects which best satisfy
these requirements.
In this model the black holes were spun up to nearly their maximal
spins during an earlier phase, when the dormant AGNs were active,
and liberate their rotational
energy in the form of UHE particles rather than in powerful jets.

The acceleration is due to the potential difference produced by
magnetic field lines threading the event horizon of the
supermassive black hole in rotation. The electromotive force
generated by a rotating black hole of mass $M$,
threaded by a magnetic field of strength $B$
extending on a range $R\approx R_g$ (gravitational radius $\equiv
G M/c^2$) is given by \cite{znajek}
\begin{equation}
emf\approx 9\times 10^{20}\left(a/M\right)B_4 M_9 \,\,{\rm volts} \,\,
\label{eq1}
\end{equation}
where $a$ is the black hole specific angular momentum,
$B_4\equiv B~/(10^4 G)$ and $M_9\equiv M/(10^9 M_{\odot})$.
For a rapidly spinning hole, we can approximate
$a/M \simeq 1$ which gives the maximal estimated energy.

Apart from the mass of the object, the only input to estimate the
maximal energy achievable in a given supermassive black hole is
the magnetic field which will depend on the mass accretion
rate. By assuming pressure equilibrium between the magnetic field
and the infalling matter we have~\cite{bg}
\begin{equation}
B_4=1.33 M_9^{-1}\dot{M}^{1/2}\,,
\label{eq2}
\end{equation}
where $\dot{M}$ is the mass accretion rate in units of $M_{\odot}/yr$.
From Eqs.~\ref{eq1} and~\ref{eq2} we obtain
\begin{equation}
emf\approx 1.2 \times 10^{21}\dot{M}^{1/2}\,\,{\rm volts}\,.\label{eq3}
\end{equation}
In the present case, we
consider protons as primary for which the energy loss dominant
processes during the acceleration phase are due to pair production
and photomeson production on ambient photons.
We will neglect these energy losses at the source because an exceptional
low accretion luminosity characterizes the supermassive black
holes at the centers of nearby bright galaxies~\cite{fabian} making
the mean free path for this interaction larger than $R_g$.
The losses due to curvature radiation induced by the magnetic
field are taken into acount as in Ref.~\cite{lev}.
Furthermore, as in~\cite{tbhl}, we assume the accretion rate onto 
the black hole to be about 10\% of the bulge mass loss rate, $\dot{M}\simeq 
0.1 M_{12}^{B}$, with $M^{B}_{12}\equiv M_{bulge}/(10^{12} M_{\odot})$,
and $M_{bulge}$ the bulge mass. The maximal proton energy expected
is then given by
\begin{eqnarray}
E_{\rm max}&=&137(\dot{M})^{1/8}(M_9)^{1/4} {\rm EeV} \nonumber\\
&\sim & 104(M^{B}_{12})^{1/8}(M_9)^{1/4} {\rm EeV}\,.\label{eq5}
\end{eqnarray}

The list of sources we will use in our numerical simulations
are given in Table~\ref{tab1} and is obtained by selecting from
the table of massive dark objects in nearby galaxies given by Magorrian
\textit{et al.}~\cite{magorrian,magorrian1},
Kormendy~\cite{kormendy}, and Marconi \textit{et
al.}~\cite{marconi} those which could accelerate particles beyond
$4\times 10^{19}$ eV. We do not include NGC 4486 as from recent
estimates of the spectral energy distribution at the core of this
galaxy it is not possible to accelerate particles to the highest
energy by the compact dynamo considered here~\cite{bg} (second
reference). Injection rates of individual sources are represented
by a weight factor given by the mass accretion rate in units of
$M_{\odot}/yr$ divided by the maximal energy in units of EeV. These
weight factors are shown in the last column of Table~\ref{tab1}.

\begin{table}
\caption{\label{tab1}{The list of sources used in our simulations. Here, $D$ is
the distance to the source in Mpc, $M_9\equiv M/(10^9 M_{\odot})$,
$M^{B}_{12}=M_{\rm bulge}/(10^{12} M_{\odot})$, $E_{max}^{19}\equiv
E_{max}/(10^{19}{\rm eV})$, $\alpha$ and $\delta$ are the
equatorial coordinates and WF is the weight factor $\times
10^{4}$}. }
\begin{indented}
\lineup
\item[]\begin{tabular}{@{}*{9}{l}}
\br
& Galaxy & D & $M_9$ & $M^{B}_{12}$ & $E_{\rm max}^{19}$ & ${\alpha}^{\circ}$ &${\delta}^{\circ}$& WF\\
\mr
1 & NGC821  &    24.1   &  0.19  &    0.12   &5.33 & 31.42 & 10.76& 2.3 \\
2 & NGC1399  &    17.9   &   5.2  &   0.32    &13.7&54.14 & -35.61 & 2.3\\
3 & IC1459  &    29.2   &   1.5   &   0.66  &11.0&343.60 & -36.29 & 6.0\\
4 &  NGC1600  &    50.2   &   11.6  &    1.29  &19.9& 67.30 & -5.19 &6.5\\
5 &  NGC2300  &    34.0   &   2.7  &    0.39    &12.0& 108.94 & 85.81&3.3\\
6 &   NGC2832 &     90.2  &    11.4  &   0.98   &19.1&139.18 & 33.96 &5.1\\
7 &   { NGC3115}  &    9.7  &   0.35  &    0.14  &6.3& 150.69& -7.48&2.2\\
8 &  { NGC3245} &     20.9  &   0.21  &    0.04  &4.7& 156.13& 28.76&0.82\\
9 & {NGC3379}&    10.6 &    0.39 &     0.076  &6.0& 161.30&12.85&1.2\\
10 &  { NGC3608}  &    22.9 &    0.25 &     0.11 & 5.6& 168.59&18.42&1.9\\
11 & { NGC4168} &    36.4 &     1.19 &    0.27  &9.3&182.43 & 13.48 &2.9\\
12 &  { NGC4261} &     31.6  &   0.52  &    0.04  &8.0& 184.21&8.02&5.6\\
13&     { NGC4278}&     17.5  &    1.56 &     0.14 & 9.1&184.41 & 29.56 &1.5 \\
14&     { NGC4291} &     26.2  &   1.86  &    0.12  &9.3& 184.52 & 75.65&1.3\\
15 &     { NGC4342}&     11.4  &   0.22  &    0.01  &4.0& 185.28& 7.32&0.25\\
16 &    { NGC4374} &     18.4  &    1.60  &    0.54  &10.8& 185.63 & 13.16&4.9\\
17 &    { NGC4459} &     16.1  &  0.07  &    0.36  &4.7& 186.62& 14.25&7.6\\
18&     { NGC4472} &     15.3  &    2.7 &     0.84  &12.9& 186.82 & 8.27&6.5\\
19&    {  NGC4473} &     15.8  &   0.34 &     0.09  &6.0& 186.82&13.71&1.6\\
20&     { NGC4486b} &     16.1  &   0.92 &    0.003  &4.9& 186.99&12.77&0.06\\
21&    {  NGC4552}&     15.3  &   0.47 &     0.14  &6.8& 188.28&12.83&2.1\\
22&     { NGC4564} &     15.3  &   0.25 &     0.04  &5.0& 188.48&11.71&0.88\\
23&    {  NGC4594} &     9.8   &  0.69  &    0.29  &8.1& 189.33&-11.33&3.6\\
24&    {  NGC4621} &     15.3  &   0.28 &    0.19  &6.1& 189.88&11.92&3.1\\
25&    {  NGC4636} &     15.3  &   0.23 &    0.32  &6.2& 190.07&2.96&5.2\\
26&    {  NGC4649} &     16.8  &    3.9 &    0.55  &13.6&190.28 & 11.83 &4.0\\
27&     { NGC4660} &     15.3  &   0.28 &    0.01  &4.4& 190.50&11.46&0.31\\
28 &   { NGC4697} &     11.7  &  0.17   &   0.20  &5.5& 191.50&-5.53&3.6\\
29&      NGC4874 &     93.3  &    20.8 &     2.08  &24.4& 194.30 & 28.23&8.5\\
30&      NGC4889 &     93.3  &    26.8 &     1.28  &24.5& 194.43 & 28.24&5.2\\
31 &   {  NGC5128} &     4.2   &  0.24   &   0.06  &5.1& 200.63&-43.76&1.1\\
32 &     NGC5252 &     96.8  &    1.0  &    0.24  &8.7& 203.93&4.80&2.7\\
33 &   { NGC5845} &     25.9  &   0.24  &    0.02  &4.4& 225.87&1.83&0.43\\
34 &    NGC6166 &     112.5 &     28.4 &     1.66  &25.7& 246.73 & 39.66&6.5\\
35 &   NGC6251 &     107.0  &   0.61  &    0.67  &8.8& 249.49&82.64&7.6\\
36 &  NGC7052 &     71.4  &   0.40  &    0.60  &7.8& 319.09&26.23&7.7\\
37 & NGC7768 &     103.1 &     9.1 &     0.89  &17.9& 357.11&26.87&5.0\\
\br
\end{tabular}
\end{indented}
\end{table}

\section{Numerical simulations}

The list of sources given in Table~\ref{tab1} has been implemented into
the numerical code for UHECR propagation in extragalactic magnetic
fields used in earlier studies, see Refs.~\cite{prev_simu,is} for details.

For each configuration many nucleon trajectories originating from
the sources were computed numerically by solving the equation of
motion for the Lorentz force and checking for pion production
every fraction of a Mpc according to the total interaction rate
with the CMB. In case of an interaction, secondary energies were
randomly selected according to the differential cross section.
Pair production by protons is treated as a continuous energy loss
process.

Each trajectory is abandoned if the particle reaches a distance from
the observer twice bigger than the distance to the farthest source,
i.e. 230 Mpc, or if the propagation time exceeds 10 Gyr.

A detection event was registered and its arrival direction and
energy recorded each time the trajectory of the propagating particle
crossed a sphere of radius 1 Mpc around the observer. For each
configuration this was done until 5000 events where registered.
For each detected event, we register also the source by which
it has been emitted.

We assume a mono-energetic injection spectrum at the source: all particles
coming from a source are emitted with the maximal energy
of acceleration for that source as given in Table~\ref{tab1}.
We take into account the individual source power by including 
the weight factor in the last column of Table~\ref{tab1}.

We assume a random turbulent magnetic field with power spectrum
$<B(k)^2>\propto k^{n_B}$ for $(2\pi/1{\rm Mpc})<k<(2\pi/0.01{\rm Mpc})$
and $\langle B(k)^2\rangle=0$ otherwise.
The magnetic field modes are computed on a linear grid in momentum
space and are Fourier
transformed onto the corresponding grid in location space.
The r.m.s. strength $B$ is given by
$B^2=\int_0^\infty\,dk\,k^2\left\langle B^2(k)\right\rangle$.
We use $n_B=-11/3$, corresponding to Kolmogorov turbulence.

This turbulent spectrum is applied to three different cases:
the first case is an all-pervading relatively weak
magnetic field $\simeq1\,$nG. For the case of a relatively strong
magnetic field we performed two simulations: in the first one the
field follows a rough representation of the Local Supercluster,
a pancake profile with scale height of 12 Mpc and scale
length of 25 Mpc; the center of the profile as at the Virgo cluster,
20 Mpc from the observer who is located in the plane of the
pancake. The maximal field strength at the Virgo cluster
equals 0.1$\mu$G. Its strength at the observer
is then $B=0.5 \times 10^{-7}\,$Gauss. In the second simulation we
consider an all-pervading field of strength $B=0.1\,\mu$G.

\section{Multi-poles and Autocorrelation function}
For each simulated sky distribution typically 1000
mock data sets consisting of $N_{\rm obs}$ observed events
were selected randomly and multiplied with the solid-angle
dependent exposure function.
For each such mock data set or for the real data
set we then obtained estimators for the spherical harmonic coefficients
$C(l)$ and the autocorrelation function $N(\theta)$. The
estimator for $C(l)$ is defined as
\begin{equation}
  C(l)=\frac{1}{2l+1}\frac{1}{{\cal N}^2}
  \sum_{m=-l}^l\left(\sum_{i=1}^{N_{\rm obs}}\frac{1}{\omega_i}Y_{lm}(u^i)
  \right)^2\,,
\label{cl}
\end{equation}
where $\omega_i$ is the total experimental exposure
at arrival direction $u^i$, ${\cal N}=\sum_{i=1}^{N_{\rm obs}}1/\omega_i$
is the sum of the weights $1/\omega_i$, and
$Y_{lm}(u^i)$ is the real-valued spherical harmonics function
taken at direction $u^i$.
For a detector at a single site we use the following parameterization:
\begin{equation}
  \omega(\delta) \propto \cos a_0\cos\delta\sin\alpha_m+
  \alpha_m\sin a_0\sin\delta\,,
\label{exposure}
\end{equation}
where $a_0$  is the latitude of the detector and $\alpha_m$ is zero for
$\xi > 1$, $\pi$ for $\xi < -1$, and $\cos^{-1}(\xi)$ otherwise,
where $\xi\equiv(\cos\theta_m-\sin a_0\sin\delta)/[\cos a_0\cos\delta]$.

In Table~\ref{tab2} we give a list of the experimental features for
the three experiments used in the next section.
In order to have maximal sky coverage for the large scale multi-poles,
we included data from the SUGAR array which operated from January 1968
to February 1979 in Australia at a latitude of $30.5^{\circ}$ South
and $149^{\circ} 38^{\prime}$ East \cite{winn}.

\begin{table}
\caption{\label{tab2}Experiments considered in the present study.
Here, ${\rm Res.}^{40}$ is the angular resolution
for energies above $40 {\rm EeV}$, $a_0^{N(S)}$ is the latitude
North (South) of the experiment and the angle $\theta_m$ is the maximal
zenith angle out to which the detector is fully efficient.}  
\begin{indented}
\lineup
\item[]\begin{tabular}{@{}*{6}{l}}
\br
Exp & ${\rm Res.}^{40}$& $a_0^N$ &  $a_0^S$ &$\theta_m$\\
\mr
AGASA& $1.6^{\circ}$ &   $+35.5^{\circ}$ &  &$45^{\circ}$ \\
SUGAR& $10^{\circ}$ &    & $-30.5^{\circ}$&$55^{\circ}$\\
AUGER& $1.^{\circ}$ &   $+35^{\circ}$ & $-39^{\circ}$&$60^{\circ}$\\
\br
\end{tabular}
\end{indented}
\end{table}

The estimator for $N(\theta)$ is defined as
\begin{equation}
N(\theta)=\frac{C}{S(\theta)}\sum_{j \neq i}
\left\{\begin{array}{ll}
1 & \mbox {if $\theta_{ij}$ is in same bin as $\theta$}\\
0 & \mbox{otherwise}
\end{array}\right\}\,,
\label{auto}
\end{equation}

where $S(\theta)$ is the solid angle size of the corresponding bin
and $C=\Omega_e/(N_{\rm obs}(N_{\rm obs}-1))$, with $\Omega_e$ denoting
the solid angle of the sky region where the experiment has
non-vanishing exposure.
In both cases for each $l$ we plot the average over all trials and realizations
as well as two error bars. The smaller error bar (shown to the left of
the average) is the statistical error, i.e. the fluctuations due to the
finite number $N_{\rm obs}$ of observed events, averaged over all realizations,
while the larger error bar (shown to the right of the average) is the
``total error'', i.e. the statistical error plus the
cosmic variance, in other words, the fluctuations due to the finite
number of events and the variation between different realizations
of the magnetic field.

In the figures shown in Sec.V, the histogram represents the data and
the solid line represents the analytical prediction for an isotropic
distribution in the case of future predictions.
Given a set of observed and simulated events we define
\begin{equation}
  \chi_n\equiv\sum_i
  \left(\frac{S_{i,{\rm data}}-\overline{S}_{i,{\rm simu}}}
             {\Delta S_{i,{\rm simu}}}\right)^n\,,\label{chi_n}
\end{equation}
where $S_{i,{\rm data}}$ refers to $C_l$ and $N(\theta)$ defined above,
obtained from the real data, and
$\overline{S}_{i,{\rm simu}}$ and $\Delta S_{i,{\rm simu}}$ are the
average and standard deviations of these quantities obtained from
the simulated data sets.
This measure of deviation from the average prediction is used to
obtain an overall likelihood ${\cal L}$ for the consistency of a
given theoretical model with an observed data
set by counting the fraction of simulated data sets with
$\chi_n$ larger than the one for the real data.
The likelihoods are computed for n=4 in Eq.~\ref{chi_n}.

\section{Results}

In the following we compare the results obtained for the
simulated UHECR propagation scenarios described above with
the observational results. As discussed in the previous section,
the comparison is based
on the statistical properties of the simulated and observed
events, expressed in terms of the angular power spectrum and
the autocorrelation function of the UHECR arrival
distributions.

\begin{figure}[h!]
\begin{center}
\includegraphics[width=0.58\textwidth,clip=true]{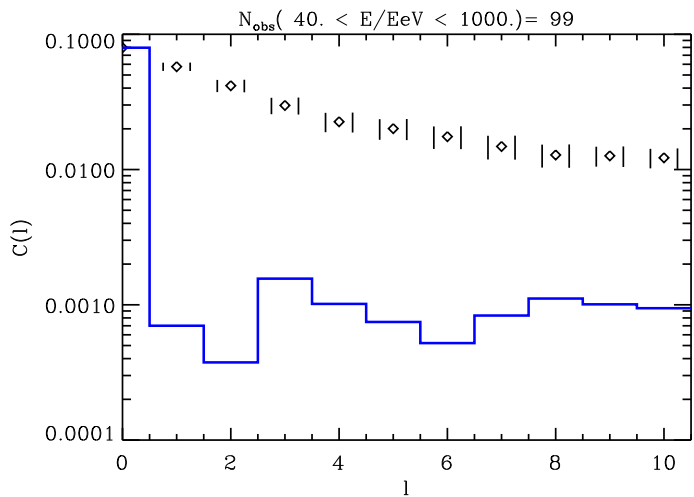}
\caption[...]{The angular power spectrum $C(l)$ as a function
of multi-pole $l$, obtained for the AGASA+SUGAR exposure function,
for ${\rm N}_{\rm obs}$=99 events observed above 40 EeV, sampled from 15
simulated  configurations with $B=1\,$nG. The diamonds indicate the
realization averages, and the left and right
error bars represent the statistical and total (including cosmic variance due
to different realizations) error, see text. The histogram represents
the AGASA+SUGAR data. The total likelihood for this fit is $\ll10^{-3}$.}
\label{F1}
\end{center}
\end{figure}

To estimate the true power spectrum from Eq.~(\ref{cl})
requires data with full sky coverage and therefore at least two
detector sites such as foreseen for the Pierre Auger experiment.
We therefore combine data from the AGASA and SUGAR experiments
which had comparable exposure in the northern and southern hemisphere,
respectively. The data set is made of 99 events: the 50 from
AGASA (excluding 7 events observed by Akeno) and the 49 from Sugar
above $4\times 10^{19}$eV. Usually care has to be taken in combining
data from two experiments with significantly different angular
resolution, see Table~\ref{tab2}. This is
possible only for multi-poles $l\le 10$ which are not sensitive to
scales $\le 10^{\circ}$, as pointed out in \cite{anch}.

In Fig.\ref{F1} we compare with the combined AGASA+SUGAR data the
angular power spectrum
predicted by simulations for a scenario with all-pervading fields
of r.m.s. strength $B=1\,$nG. The large
predicted anisotropy is due to the fact that the sources follow
the supergalactic plane and that UHECRs do not diffuse on the
scale of the distance to the dominant, relatively close sources, 
since their Larmor radius
$r_{L}\simeq110(E/10^{20}eV)(B/10^{-9}G)^{-1}\,$Mpc. In addition,
the distribution of events is concentrated around the directions
to a few close sources which give the principal contribution to
the detected flux. This can be understood by considering, in
Fig.\ref{F2}, the number of events detected from source ``$i$'' in
a given range of energy $n_i(E_{1}<E< E_{2})$, normalized to the
total number of events detected from all sources in the same
energy range, $n_{\rm tot}(E_{1}<E< E_{2})$, for this weak field
case. The labels on the x axis correspond to the distances of
sources in Mpc whereas at the top of each bin we mark the maximal
acceleration energy of that source. Almost all particles are
detected in the  range between 40-100 EeV, whereas only $12\%$ of the
total 75000 particles are detected outside this interval.
The corresponding histogram also shows that only the
sources in the list within 20 Mpc from Earth give a significant
contribution. In case of negligible deflection we expect that clusters
just reflect the point-like sources; if the number of observed
events $N_{\rm obs}$ is larger than the number of contributing sources,
as in the present scenario, each source contributes on average
more than one event and strong clustering is expected. In addition,
most of the closest sources are within an angular distance less than
$5^{\circ}$ from each other. This implies that most of the events
come from the same direction, and leads to a predicted
autocorrelation function which at scales of a few degrees is about
30 times the value obtained from AGASA data.

The contribution of sources injecting above 40 EeV is strongly suppressed
below this energy by two effects: the closest sources can give a
contribution at low energies only when their maximal acceleration
energy is close to 40 EeV, but they are sparse in the list considered
here; the source NGC 4342 at 11.4 Mpc with ${\rm E}_{\rm
max}=40\,$EeV is responsible for the first peak in the distribution
below 40 EeV in Fig.~\ref{F2}.
The farthest sources in contrast do not contribute below
$\simeq40\,$EeV because their maximal acceleration energy is
sufficient to keep the energy of the particles above 40 EeV in the
non diffusive regime. Fig.~\ref{F2} shows that the contribution
from the farthest sources is actually negligible at all energies,
due to suppression by a factor proportional to ``weight
factor$/D^2$'', where $D$ is the distance to the source.

\begin{figure}[h!]
\begin{center}
\includegraphics[width=0.45\textwidth,clip=true,angle=270]{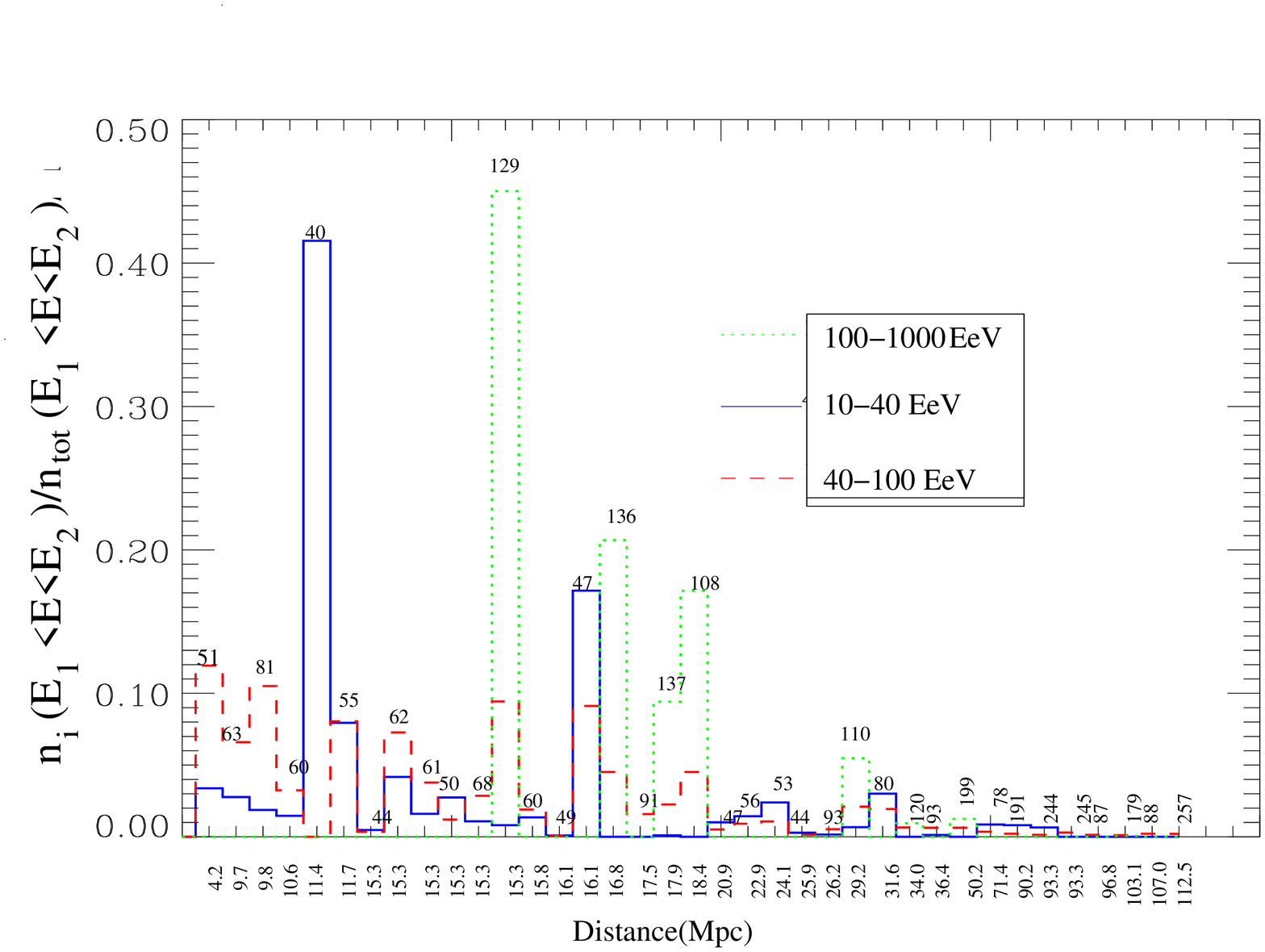}
\caption[...]{The number of detected events per source
in a given range of energy, normalized to the total number of events
detected in the same range of energy, averaged over 15 realizations,
for the simulation with $B=1\,$nG. The x axis represents the
distance corresponding to each source in Mpc.
At the top of each histogram the maximal acceleration energy
per source is shown. The solid line corresponds to the range (10-40) EeV, the
dashed one to the range (40-100) EeV, and the pointed one to the range
(100-1000) EeV.}
\label{F2}
\end{center}
\end{figure}

So far we have found that the list of sources, as given in
Table~\ref{tab1}, cannot reproduce current data. If typical deflections
are small, consistency with the data requires source distributions
that are more homogeneously distributed than our sample in
Table~\ref{tab1}. If at least some of the sources in our list
contribute significantly to the total flux, we would still expect
a significant correlation between the arrival direction of the events and the
positions of some of the objects in our list.

Following the same approach as in Ref.~\cite{lev} we made a simple
estimate of the positional coincidence between the sources in
Table~\ref{tab1} and the UHECR events observed by AGASA with energy
above $4\times 10^{19}$ eV. We calculate the number of real
coincidences in circles centered on the position of a given source.
For a circle radius of $4.8^\circ$, corresponding to a  $3\sigma$
error box in angular resolution, coincidences with the AGASA data
were only found for two relatively far sources, namely NGC4874 and
NGC4889, and the source NGC 821 at 24 Mpc, 
which contribute negligibly to the flux, see Fig.~\ref{F2}.
The source with the largest individual contribution (NGC5128)
would predict $\simeq6$ events in the range between 40 and
100 EeV, with a Poisson probability for no coincidence of about 0.1\%.
This implies that either deflection has to be significant and/or a
considerable part of the observed flux is due to sources not in
our list. We note in this context that hints of UHECR
arrival direction correlations with another list of quasar remnants
have been reported in Ref.~\cite{tbhl}.

We now investigate whether stronger magnetic fields, by providing
larger angular deflection, might provide a better match to the
observational data for this same source distribution. We performed
simulations in turbulent magnetic fields following the profile of a
sheet centered at 20 Mpc from Earth with strength $0.1 \mu$G at
the center of the sheet. We recall that infinitely extended profiles
are consistent with Faraday rotation limits for fields up to fractions
of a micro Gauss~\cite{blasi}. In Figs.~\ref{F3} and~\ref{F4}
we show the autocorrelation function and angular distribution,
respectively, above $4\times10^{19}\,$eV, predicted by a simulation
performed with fields following a pancake profile of scale length
25 Mpc and scale height 12 Mpc. The predictions are still not consistent
with the data for our set of sources. Scenarios for pancake scale
heights up to $\simeq25\,$Mpc lead to similar results.

\begin{figure}[h!]
\begin{center}
\includegraphics[width=0.58\textwidth,clip=true]{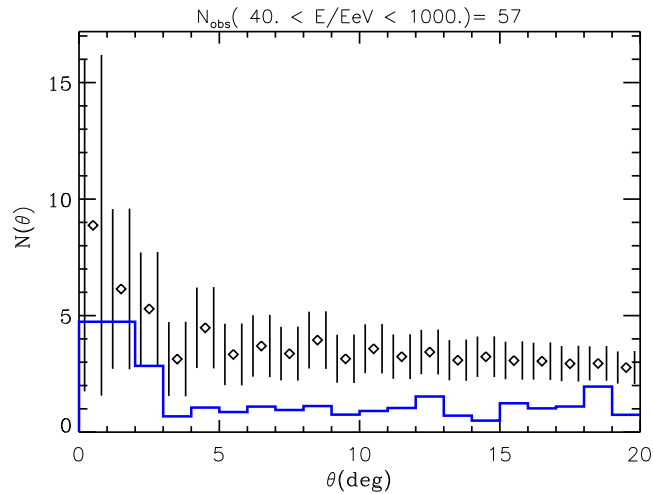}
\caption[...]{The angular correlation function $N(\theta)$, eq.~(\ref{auto}),
as a function of the angular distance $\theta$, using a bin size
of $\Delta\theta =1^{\circ}$, obtained for the AGASA exposure
function, for ${\rm N}_{obs}$=57 events observed above 40 EeV, sampled from
10 simulated realizations for a magnetic field following a pancake
profile of scale length 25 Mpc and scale height 12 Mpc with
$B=0.1\mu$G at the center.
The diamonds indicate the realization averages, and the left and
right error bars represent the statistical and total (including
cosmic variance due to different realizations) errors respectively;
see text for explanations. The histogram represents the AGASA data.
The total likelihood for this fit is 0.01.}
\label{F3}
\end{center}
\end{figure}

\begin{figure}[h!]
\begin{center}
\includegraphics[width=0.58\textwidth,clip=true]{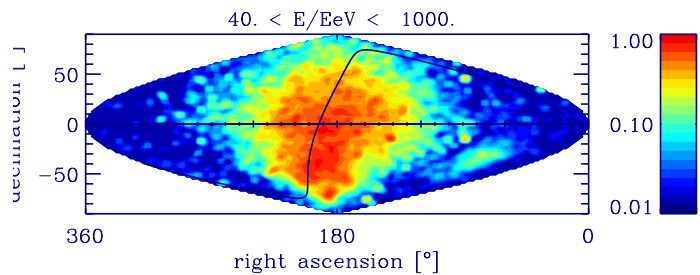}
\caption[...]{Angular distribution in equatorial coordinates on the
celestial sphere for the same scenario as in Fig.~\ref{F3}.
The solid line marks the supergalactic plane. The grey scale represents the
integral flux per solid angle. The pixel size is $1^\circ$; the image has
been convolved to an angular resolution of 1.6$^\circ$ corresponding to
AGASA.}
\label{F4}
\end{center}
\end{figure}

In the scenario discussed above, the profile of the magnetic
field in the sheet is such that the field is stronger in the
middle plane than on the boundaries of the sheet. In Figs.~\ref{F5}
and~\ref{F6} we show the angular distribution and the
autocorrelation function, respectively, in the case where a
magnetic field is distributed homogeneously rather than
following a profile, although this is not a realistic scenario
for our structured universe.
As a result, when the effects of the boundary are absent, the
predictions are more consistent with the data and the final distribution
appears more isotropic.

\begin{figure}[h!]
\begin{center}
\includegraphics[width=0.58\textwidth,clip=true]{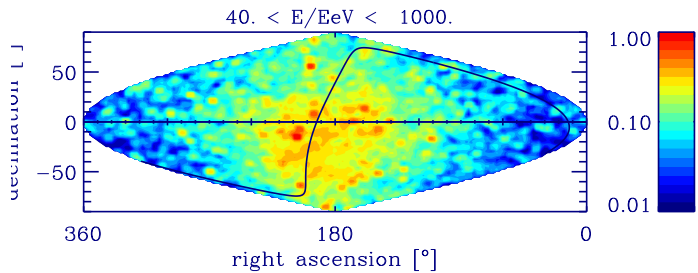}
\caption[...]{Same as Fig.\ref{F4}, but for a magnetic field
$B=0.1 \mu$G distributed homogeneously.}
\label{F5}
\end{center}
\end{figure}

\begin{figure}[h!]
\begin{center}
\includegraphics[width=0.58\textwidth,clip=true]{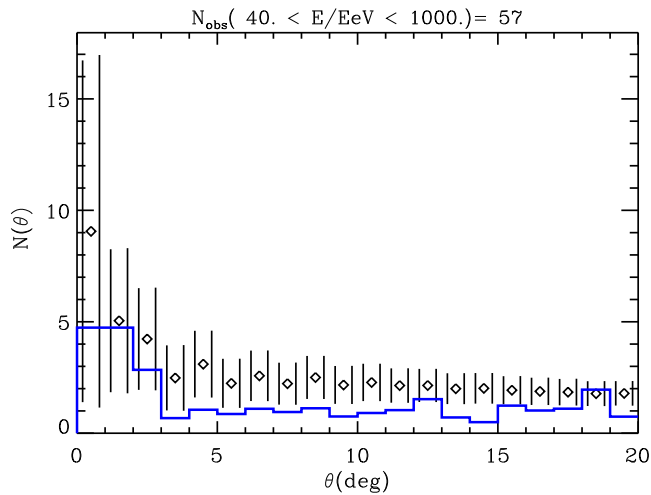}
\caption[...]{Same as Fig.~\ref{F3}, but for homogeneous field
of $0.1\mu$G. The total likelihood for this fit is 0.18.}
\label{F6}
\end{center}
\end{figure}

\begin{figure}[h!]
\begin{center}
\includegraphics[width=0.58\textwidth,clip=true]{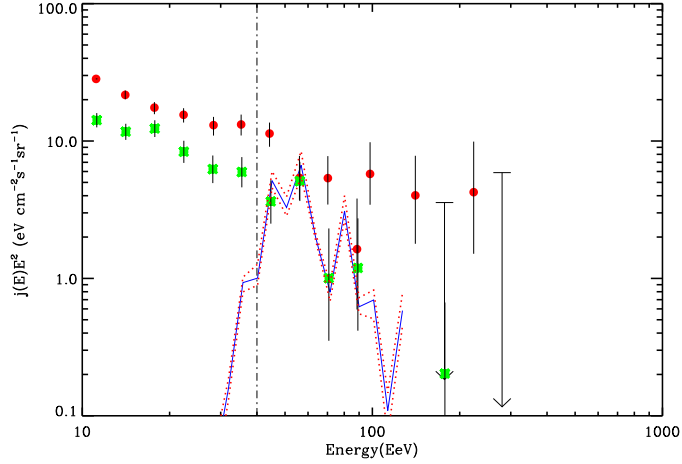}
\caption[...]{The realization averaged energy spectra
corresponding to Figs.~\ref{F5},\ref{F6}.
The solid line represents the spectrum that
would have been detected by AGASA, using the exposure
Eq.~(\ref{exposure}). The dotted lines are the fluctuations
due to the different magnetic field realizations and
the one sigma error bars indicate the AGASA data
(full dot) and HiRes data (asterisk).}
\label{F7}
\end{center}
\end{figure}

In Fig.~\ref{F7} we compare the spectrum predicted for this
scenario with the AGASA and HiRes data.
At the highest energies the cut-off is simply due to the fact
that the maximal injection energy is 257 EeV, whereas we did not
attempt to explain the flux below $\sim40\,$EeV as we only included
sources accelerating beyond this energy in our simulations.
Less massive quasar remnants could well significantly contribute
in this energy range as well.

The spectrum has been normalized to optimally fit the
AGASA data which requires an average UHECR emission power of
$\sim10^{41}$erg/s for the nearby sources. At the same time the
maximum power that can be extracted from a Kerr black hole is given by
$L_{BH} \simeq 10^{40}(a/M)^2 M_6^2B_4^2
\,\,{\rm erg}\,\, {\rm s}^{-1}$~\cite{levboldt}. Using Eq.~(\ref{eq2}),
the definition of $\dot{M}$ and the approximation
$(a/M)\simeq 1$, as before, we parametrize the UHECR power as
\begin{equation}
L_{CR}\simeq \alpha_{CR} (0.17) 10^{46} {M}_{12}^B\,\,{\rm erg}
\,{\rm s}^{-1}\,,
\label{luminosity}
\end{equation}
where $\alpha_{CR}$ is the fraction of the power emitted as UHECRs.
Comparing these two numbers and using typical bulge masses given
in Table~\ref{tab1} for the closest sources implies a required
UHECR injection efficiency of about $1\%$ of the accretion rate.

The spectrum in Fig.~\ref{F7} shows no very pronounced GZK cut-off.
This provides an important test for future
experiments, especially the Pierre Auger project, which will be
able to eventually confirm or falsify the presence of a GZK cut-off
in the spectrum. If this cut-off is not observed, the quasar remnant
scenario for UHECR origin here investigated could be promising
good possibility but with a larger sample of sources such as to
reproduce the observed  isotropy at large scale. Future projects
may provide sufficient statistics to probe the wiggles seen in
Fig.~\ref{F7} predicted by the monoenergetic injection.

In Fig.\ref{F8} we shown the angular power spectrum for the
exposure of the Pierre Auger experiment with full sky coverage,
assuming 1500 events observed
above 40 EeV, for the case of an all-pervading $0.1\,\mu$G field.
For the exposure function we add Eq.~\ref{exposure} for two sites
located at $a_0=-35^{\circ}$ and at $a_0=39^{\circ}$.
The solid line represents the analytical prediction for a fully
isotropic distribution, predicting $C_l\simeq(4\pi N_{\rm obs})^{-1}$.
This scenario predicts an anisotropy that should be easily
detectable by the Pierre Auger experiment.

\begin{figure}[h!]
\begin{center}
\includegraphics[width=0.58\textwidth,clip=true]{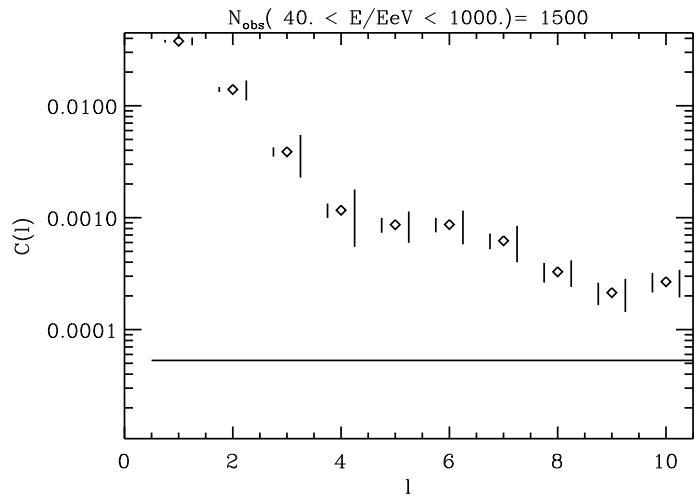}
\caption[...]{The angular power spectrum $C(l)$ as a function of
multi-pole $l$, obtained for the two-site Auger exposure function,
assuming ${\rm N_{\rm obs}}$=1500 events observed above 40 EeV, sampled from 10
simulated realizations for an all-pervading turbulent $B=0.1\,\mu$G field.
The solid line represents the analytical prediction for an
isotropic distribution.}
\label{F8}
\end{center}
\end{figure}

\section{Conclusions}

In the present work we considered a model where the
sources of ultra-high energy cosmic rays are quasar
remnants or dormant AGNs, with underlying supermassive black holes
as suggested in Ref.~\cite{bg}. We assumed a list of 37 of such
objects injecting ultra high energy cosmic rays at an energy
and with a power determined by their mass and accretion properties. We then
studied the effects of propagation in different extragalactic
magnetic fields scenarios on predicted distributions of arrival
energies and directions.
As statistical quantities for this analysis we used spherical
multi-poles and the autocorrelation function.
We found that for a weak magnetic field, of order of $1\,$nG,
the predictions appear to be inconsistent with
the observed distribution, as already pointed out in Ref.~\cite{is},
because the magnetic field is too weak to isotropize the distribution
coming from a limited number of non-uniformly distributed
sources, as in our list.

We also found that the contribution from the farthest sources
is completely negligible even for this weak magnetic field;
this is true to an even larger degree for stronger fields,
consistent with Ref.~\cite{is}.

We found no coincidences between the position of the sources in
our list and the arrival direction of the AGASA events. In the
scenario with a weak magnetic field, we thus conclude that the list
of quasar remnants considered here can not provide the
dominant contribution to the observed flux. However, hints
of correlations with another set of objects subjected to contain
quasar remnants have been reported elsewhere~\cite{tbhl}.

These results show that if quasar remnants are the sources of
ultra high energy cosmic rays, many more of them than considered here
must contribute and/or an extended extragalactic magnetic field of
$\sim0.1\,\mu$G must exist.
We note, however, that the number of sources cannot
be arbitrarily high, as it would get in conflict with the estimates
of total density and mass function of supermassive black holes
at the current epoch~\cite{marconi}.

We also performed simulations
with magnetic fields of $\sim0.1\,\mu$G following pancake profiles
of various extensions. These two-dimensional sheets
were centered at the Virgo cluster, at about 20 Mpc from
the observer. For a sheet dimension of $\simeq20\,$Mpc in the
direction towards the observer, and $\simeq10-20\,$Mpc perpendicular
to it, the predicted large scale anisotropy and the autocorrelation
beyond a few degrees are too large to be consistent with the data.
The arrival directions are concentrated along the supergalactic plane
and the conclusions are basically the same as for the weak field
case. Only for all-pervading fields of $\sim0.1\,\mu$G with
homogeneous properties the predictions become more consistent with
the data. However, this is unlikely to represent a realistic
description for our structured universe.

The quasar remnant scenario tends to predict a relatively hard
spectrum with at best a mild GZK cut-off which
leaves the possibility for this kind of objects to be
sources at the highest energies. Nevertheless, we recall that
these results could be affected by the assumption of a mono-energetic
injection, which is the more optimistic
case but not necessarily the most realistic. Furthermore, if the flux at
highest energies is dominated by just a few sources, mono-energetic
or very hard injection spectra may lead to conspicuous wiggles
in the spectrum. Finally, rough estimates show that about 1\%
of the accretion rate emitted in ultra-high energy cosmic rays
suffices to reproduce the observed flux level.

As already pointed out in Ref.~\cite{lev}, this scenario also has
some direct observational consequences at much lower energies which
could be tested in the near future.
The emitted spectrum of curvature photons peaks in the
TeV band; therefore, these sources could be detected by future
experiments like GLAST and MAGIC~\cite{lev}.
At the highest energies, the present development of large new detectors,
such as the Pierre Auger~\cite{auger}
experiment and the space-based air shower detectors such as  OWL~\cite{OWL}
and EUSO~\cite{EUSO}, will considerably
increase statistics. All of them are planned to achieve full sky coverage
and anisotropies predicted by the quasar remnant scenario should
be easily detectable.

\ack
GS would like to thank Martin Lemoine, Francesco Miniati, and
Torsten En{\ss}lin for earlier collaborations on the codes partly
used in this work. CI would like to thank Martin Lemoine, Peter Biermann
and John Kormendy for useful comments. We thank
Joe Silk for previous collaboration and discussion at an early stage
of this work. CI is partially supported by MIUR under COFIN PRIN-2001.
GB was supported by the DOE and the NASA grant NAG 5-10842 at Fermilab.

\end{document}